\documentstyle[psfig]{lamuphys}
\newcommand{\ncm}{\newcommand}
\ncm{\BE}{\begin{equation}}   \ncm{\EE}{\end{equation}}
\ncm{\BDM}{\begin{displaymath}} \ncm{\EDM}{\end{displaymath}}
\ncm{\BEQ}{\begin{displaymath}} \ncm{\EEQ}{\end{displaymath}}
\ncm{\BAR}{\begin{array}}     \ncm{\EAR}{\end{array}}
\ncm{\BEA}{\begin{eqnarray}}  \ncm{\EEA}{\end{eqnarray}}
\ncm{\tf}{\textstyle\frac}
\ncm{\vpd}{\frac{\varphi}{3}} \ncm{\vqd}{\frac{\pi-\varphi}{3}}
\ncm{\del}{\delta}            \ncm{\hal}{{\tf{1}{2}}}
\ncm{\lra}{\leftrightarrow}   \ncm{\Wb}{{\bar{W}}}
\ncm{\ra}{\rightarrow}        \ncm{\AO}{{\cal A}}
\ncm{\lam}{\lambda}           \ncm{\sig}{\sigma}
\ncm{\ep}{\epsilon}           \ncm{\Ga}{\Gamma}
\ncm{\al}{\alpha}             \ncm{\WB}{\bar{W}}
\ncm{\om}{\omega}             \ncm{\TT}{\bar{T}}
\ncm{\Lb}{\left[}             \ncm{\Rb}{\right]}
\ncm{\lb}{\left\{}            \ncm{\rb}{\right\}}
\ncm{\lk}{\left(}             \ncm{\rk}{\right)}
\ncm{\HH}{{\cal H}}           \ncm{\AAA}{{\cal A}}
\ncm{\PP}{{\cal P}}           \ncm{\BB}{{\cal B}}
\ncm{\GG}{{\cal G}}           \ncm{\JJ}{{\cal J}}
\ncm{\hx}{\hspace*{2mm}}      \ncm{\hi}{\hspace*{12mm}}
\ncm{\hs}{\hspace*{1cm}}      \ncm{\su}{\sum_{j=1}^L}
\ncm{\hq}{\hspace*{6mm}}      \ncm{\ny}{\nonumber}
\ncm{\th}{\theta}             \ncm{\NIF}{N\rightarrow\infty}
\ncm{\nh}{\frac{N}{2}}        \ncm{\vph}{\varphi}
\ncm{\epf}{\frac{\pi}{5}}     \ncm{\vpf}{\frac{4\pi}{5}}
\ncm{\zpf}{\frac{2\pi}{5}}    \ncm{\ZZ}{\hbox{Z\hspace{-4pt}Z}}
\ncm{\tepf}{{\textstyle\epf}} \ncm{\sib}{\mbox{sin}}
\ncm{\sbq}{\sib^2\beta}       \ncm{\cbq}{\mbox{cos}^2\beta}
\ncm{\spf}{\sin{\tepf}}       \ncm{\cpf}{\cos{\tepf}}
\ncm{\pia}{{\tf{\pi}{8}}}     \ncm{\pivi}{{\tf{\pi}{4}}}
\ncm{\pis}{{\tf{\pi}{6}}}     \ncm{\wud}{\sqrt{3}}
\ncm{\piz}{{\tf{\pi}{12}}}    \ncm{\tib}{\mbox{tan}^2} 
\ncm{\RIF}{R\rightarrow\infty} \ncm{\pid}{{\tf{\pi}{3}}}
\ncm{\spc}{\\[4mm]}            \ncm{\prj}{\prod_{j=1}^{m_P}}
\ncm{\prl}{\prod_{l=1}^{m_P}}  \ncm{\mo}{\;\mbox{mod}\;3} 
\begin{document}
\title{Finite-size energy levels of the superintegrable chiral Potts 
model}
\author{G.\,\,von\,\,Gehlen}
\institute{Physikalisches Institut der Universit\"at Bonn, Nussallee 12,
53115 Bonn, Germany}
\maketitle
\begin{abstract} In the solution of the superintegrable chiral Potts model
special polynomials related to the representation theory of the Onsager 
algebra play a central role. We derive approximate analytic formulae for 
the zeros of particular polynomials which determine sets of low-lying energy 
eigenvalues of the chiral Potts quantum chain. These formulae allow 
the analytic calculation of the leading finite-size corrections to the 
energy eigenvalues without resorting to a numerical determination of the zeros. 
\end{abstract}
\section{Introduction}
The chiral $Z_N$-symmetrical Potts model, apart from showing a rich phase
structure, has attracted considerable interest because for appropriate
parameter choi\-ces it has very special integrability properties. In the
"superintegrable" case it is integrable {\it both} because it provides a
representation of Onsager's algebra (in this respect being a natural
generalization of the Ising model), and also because its Boltzmann
weights satisfy a new type of Yang-Baxter equation. Both properties
guarantee the existence of an infinite set of commuting charges. In order 
to calculate the spectrum, functional relations are
needed. Such relations had first been conjectured by \cite{AMCP}, and later
derived by exploring relations to the six-vertex model at
$q=e^{2\pi i/N}$ (\cite{BS}, \cite{BBP}). The formulae which give the 
spectrum in the superintegrable case have been given by 
\cite{AMCP} for $Z_3$ and for general $Z_N$ by \cite{Baxsc}. 
Little is known about correlation functions. 
For the general integrable case (not satisfying the Onsager algebra), 
the free energy and interface tensions have been obtained in the 
thermodynamic limit (\cite{ORB}). 
It is still a challenge to find an analytic derivation of the order 
parameter (\cite{Bax98}). Recently, general integrable boundary conditions 
for the chiral Potts model have been obtained by \cite{YKZ}. 
\par In the following we concentrate on the solution of the superintegrable 
case, in which special polynomials related to the representation theory of 
the Onsager algebra play a central role. The energy eigenvalues are 
determined in terms of the zeros of these polynomials. For the lowest 
states the polynomials are of order $m_E=\Lb((N-1)L-Q)/N\Rb$. 
$L$ denotes the number of sites of 
the chain and $Q$ the $Z_N$-charge sector. $\Lb x\Rb$ stands for the
integer part of a rational number $x$. In the limit $L\ra\infty$ the sum
over the terms depending on the zeros can be expressed as a contour integral 
without explicitly calculating the zeros. This method has been used to obtain
the ground state energy and excited levels in the thermodynamic limit. 
For the gapless regime finite-size corrections have been studied by \cite{AC}
in order to obtain information on correlation 
functions through arguments of conformal invariance. Interesting effects
due to the broken parity invariance have emerged.    
For finite $L$ one has to resort to numerical methods. 
\par In these lecture notes, after reviewing main features of the model 
and its solution, we study the zeros of the above mentioned polynomials. 
We derive quite accurate approximate analytic expressions for their 
location at finite $L$. 
\section{Integrability of the chiral Potts model}
\subsection{The integrable model}
For many discussions of the chiral Potts quantum chain, it turns out to
be sufficient to consider the following $Z_N$-symmetric hamiltonians which 
depend on three real parameters $\lam,\;\phi,\;\vph$:
\BEQ H=-\su\sum_{l=1}^{N-1}\frac{1}{\sin{(\pi l/N)}}
  \left(\;e^{i\vph(2l-N)/N}\;Z_{j}^l 
  +\lam\;e^{i\phi(2l-N)/N}\;X_{j}^l \,X_{j+1}^{N-l}\right).  \EEQ
The first sum is over the sites $j$. At each site $j$ there are operators
$X_j$ and $Z_j$ acting in a vector space $C^N$  and satisfying
\BEQ Z_i X_j=X_j\:Z_i\:\om^{\delta_{i,j}};\hi Z_j^N= X_j^N=1;\hi
 \om=e^{2\pi i/N}.  \label{sig}\EEQ
Often it is useful to represent the operators $X_j$ and $Z_j$ by
$(X_j)_{l,m}=\delta_{l,m+1}$ $(mod\;N)$ and 
$Z_j=diag(1,\om,\om^2,\ldots,\om^{N-1})_j.$ 
$H$ commutes with the $Z_N$-charge operator $\hat{Q}=\prod_{j=1}^L Z_j$.
We write the eigenvalues of $\hat{Q}$ as $\om^Q$ where
$\;Q=0,1,\ldots,N-1$. $\lam$ is the inverse temperature. We shall consider 
the periodic case $X_{N+1}=X_1$, although, as shown by \cite{Baxsc}, twisted 
toroidal boundary conditions require little additional effort. 
\subsection{Particular cases: Ising, Parafermionic} 
For $N=2$ the sum over $l$ has only a
single term. The angles $\phi,\;\vph$ drop out, $Z_j$ and $X_j$
become Pauli matrices and we obtain the Ising quantum chain. If we put
$\phi=\vph=0$ then $H$ describes the Fateev-Zamolodchikov parity invariant
parafermionic $Z_N$-quantum chain. For $N\ge 3$ and  $\phi\neq 0$ or
$\vph\neq 0$ parity is broken (therefore $\phi$ and $\vph$ are called
chiral angles) and this gives rise to several interesting features of the
model, e.g. to the appearance of incommensurate phases and oscillating 
correlation functions.
\subsection{Yang-Baxter-integrable case} 
If $\cos{\vph}=\lam\cos{\phi}$ then $H$ can be derived 
from a two-dimensional lattice model defined by the following transfer 
matrix which depends on rapidities $p$ and $q$:
\BE    T_{p,q}(\{l\},\{l'\})\;=\;\prod_{j=1}^L\;
W_{p,q}(l_j-{l'}_j)\;\WB_{p,q}(l_j-{l'}_{j+1})\label{tr} \EE
$\{l\}$ and $\{l'\}$ are the sets of $N$-valued spin variables at
alternating vertex rows of the diagonally drawn lattice. The rapidity
lines $p$ run horizontally on the dual lattice, the $q$ vertically.  
The Boltzmann weights $W$ and $\WB$ are defined in terms of the functions
$x_p,\,y_p,\,\mu_p;\;x_q,\,y_q,\,\mu_q$ of the rapidities $p$ and $q$
by (\cite{AuY})\BEQ W_{p,q}(n)=\lk\frac{\mu_p}{\mu_q}\rk^n
\prod_{j=1}^n\frac{y_q-\om^j x_p}{y_p-\om^j x_q};\hs
 \Wb_{p,q}(n)=\lk\mu_p\mu_q\rk^n \prod_{j=1}^n\frac{\om x_p-\om^j x_q}
 {y_q-\om^j y_p}   \EEQ   
where $n=0,\;1,\;\ldots,\;N-1$. The requirement of
$Z_N$-symmetry imposes several restrictions on the six functions, e.g.
\BEQ \lam(x_q^N y_q^N+1)=x_q^N+y_q^N. \EEQ
where $\lam$ is a parameter describing the inverse temperature. At fixed
$\lam$ only one
of the three functions $x_q,\,y_q,\,\mu_q$ is independent, the other two 
 are determined by nonlinear relations.
\par In the Ising case $N=2$, the functions $x_q$ etc. can be simply
expressed in terms of the meromorphic
elliptic functions of modulus $\lam$:
\BEQ x_q=-\sqrt{\lam}\;\mbox{sn}\,q;\hs
  y_q=\sqrt{\lam}\; \mbox{cn}\,q/\mbox{dn}\,q,\hs
     \mu_q=\sqrt{1-\lam^2}/\mbox{dn}\,q\EEQ and one obtains the
parametrization of the Ising Boltzmann weights known e.g. from \cite{Baxb}. 
Integrability is guaranteed by the following Yang-Baxter-equation,
which contains the $p$- and $q$-variables {\it separately}  \BEA
\lefteqn{\sum_{l=0}^{N-1}W_{p,q}(l'-l)\Wb_{r,q}(l''-l)W_{r,p}(l-l''')}
 \ny\\&&\hs =\;
  R_{pqr}\Wb_{r,p}(l''-l')W_{r,q}(l'-l''')\Wb_{p,q}(l''-l'''). \ny \EEA
The explicit expression for $R_{pqr}$ can be found e.g. in \cite{AMCP}.
The superintegrable case corresponds to fixing the $p$-dependent
functions to $x_p=y_p$ and $\mu_p=1$ so that we are left with only one
independent function, e.g. $x_q$. \\ $H$ is obtained if at fixed $p$ we
expand the transfer matrix for small $q-p$.
\subsection{The superintegrable case, Onsager's algebra}
The choice $\:\phi=\vph=\pi/2\:$ gives the "superintegrable" case of the
hamiltonian. Writing $\;\;H=\hal N\,(A_0+\lam A_1)\;\;$ we have
\begin{equation} 
 A_0=-\frac{4}{N}\su\sum_{l=1}^{N-1}\frac{Z_j^l}{1-\om^{-l}};
      \hs A_1=-\frac{4}{N}\su\sum_{l=1}^{N-1}\frac{X_j^l
                   X_{j+1}^{N-l}}{1-\om^{-l}}.\label{hsup}\end{equation}  
It has been shown by \cite{GR} that $A_0$ and $A_1$ satisfy the \cite{DG} 
conditions  \BE \Lb A_0,\Lb A_0,\Lb A_0,A_1\Rb\,\Rb\,\Rb\,=\,16\Lb A_0,A_1\Rb;
 \hx\; \Lb A_1,\Lb A_1,\Lb A_1,A_0\Rb\,\Rb\,\Rb\,=\,16\Lb A_1,A_0\Rb.
\label{DG} \EE
Due to these conditions, $A_0$ and $A_1$ generate Onsager's
algebra $\AO$:  \BEQ   [A_l,A_m] = 4\,G_{l-m} \EEQ
\BEQ [G_l,A_m]= 2\,A_{m+l}-2\,A_{m-l};\hs  [G_l,G_m]= 0;\hs
 l,\:m\in \ZZ. \EEQ
which has been an essential tool in the \cite{Ons} original solution of the
Ising model. $\AO$ implies an infinite set of constraints:
\BEQ [A_2,A_1]=[A_1,A_0]=[A_0,A_{-1}]=\ldots \EEQ and the existence of the
infinite set of commuting charges $Q_m$:
\BEQ Q_m=\hal\lk A_m+A_{-m}+\lam(A_{m+1}+A_{-m+1})\rk
\hs\mbox{with}\hs  Q_0=H. \EEQ 
Finite dimensional representations of $\AO$ are obtained if there is a 
recurrence relation of finite length among the $A_l$ 
(Davies 1990, Roan 1991): \BEQ
\sum_{k=-n}^n\;\alpha_k\,A_{k-n}\;=0.  \EEQ 
If $z_j$ ($j=1,\ldots,m_E$) are the zeros of the polynomial 
$f(z)=\sum_{k=-n}^n\alpha_k z^{k+n}$, then we can express 
the $A_m$ and $G_m$ in terms of a set of $sl(2,C)$-generators
\BEQ A_m=2\sum_{j=1}^n\:\lk z_j^m\,E_j^+ +z_j^{-m}
 E_j^-\rk;\hs G_m\:=\:\sum_{j=1}^n\:\lk z_j^m-z_j^{-m} \rk \:H_j \EEQ
where \BEQ [E_j^+,E_k^-]\:=\:\delta_{jk}\:H_k; \hs\hx
    [H_j,\:E_k^\pm]\:=\:\pm 2\,\delta_{jk}\:E_k^\pm. \EEQ
So, $\AO$ is isomorphic to a subalgebra of the loop algebra of a direct sum
of $sl(2,C)$ algebras. 
\par Each sector of the model is characterized by a set
of points $z_j$, which, of course, are not fixed by the algebra $\AO$
alone. For the hamiltonian (\ref{hsup}) the polynomials $f(z)$, or rather 
related polynomials $\PP_Q(s)$, have been obtained by \cite{AMCP} 
and \cite{Baxsc} from functional equations for the transfer matrix (\ref{tr})
of the two-dimensional model, as will be reviewed in the next section. 
\\As B. \cite{Dav} has shown, from the property (\ref{DG}) of $H$ it follows
that {\it all} eigenvalues $E$ of $H$ depend on $\lam$ in the form 
\begin{equation} E\,=\,a\,+\,b\,\lam\,+\,N\sum_{j=1}^{m_E}\pm
 \sqrt{1+2\lam\cos{\th_j}+\lam^2} \label{EE}\end{equation} 
where $a$ and $b$ are integers and 
$\;\;\cos{\th_j}=\hal\lk
  z_j+z_j^{-1}\rk.\;\;$ So the main task of the
diagonalization of the superintegrable $H$ reduces to the
calculation of the $\cos{\th_j}$ or equivalently, of the zeros of 
the polynomials $f(z)$. For hermitian hamiltonians, the $z_j$ must be on
the unit circle. 

\section{Solution of the $Z_3$-superintegrable model}
\subsection{Functional equation and ansatz of Albertini et al.}
\cite{AMCP} discovered the following functional
equation for the $Z_3$ superintegrable case with the $p$-variables fixed
to $x_p=y_p=\lk\frac{1-\lam}{1+\lam}\rk^{1/6}$: Splitting off some
poles from $T_{p,q}$ they get a meromorphic function $\TT_q$ which 
satisfies \BEA \lefteqn{ \TT_q\:\TT_{Rq}\:\TT_{R^2q}\;=\;3^L e^{-iP}\:\lb\;
   (t-1)^L(\om^2 t-1)^L\TT_q\right.}\ny \\ &&\left.\hs
   + (\om t-1)^L(\om^2 t-1)^L\TT_{R^2q}
   + (t-1)^L\;(\om t-1)^L\;\TT_{R^4q}\;\rb  \label{fff} \EEA
where $P$ is the momentum operator, $t=x_qy_q/x_p^2$ and $R$ is the 
mapping\\ 
$R(x_q,y_q,\mu_q)=(y_q,\om x_q,\mu_q^{-1})$. Solutions to this functional
equation are then obtained from the ansatz  \BE \TT_q=\;\;\ldots\;\;
\prod_{l=1}^{m_P}\lk\frac{1+\om  v_l t}{1+\om v_l}\rk\prod_{j=1}^{m_E}
\lb(1-\lam)(\mu_q^3+1)\pm 2 w_j(\mu_q^3-1)\rb.  \label{ans}\EE 
We skip here writing some factors relevant for the determination of
the linear terms in (\ref{EE}) and chose a certain integer $P_b=0$: 
for details see \cite{AC} or Baxter (1993,1994). The ansatz (\ref{ans}) 
contains $m_P$ excitations with
Bethe-type rapidities $v_l$, and in the last factor, $m_E$ functions $w_j$. 
It requires some calculation to see that if we put  \BEQ
w_j=\hal \sqrt{1+2\lam\frac{1+t_j^3}{1-t_j^3}+\lam^2} \EEQ
then the left hand side of (\ref{fff}) vanishes, and in order to get the
right hand side vanishing too, $t_j$ must be zero
of the polynomial  \BEA \lefteqn{\PP_Q(t^3)\;=\;\frac{t^{-c}}{3}\lb
  (t^2+t+1)^L\prod_{l=1}^{m_P}\frac{1+v_l t}{1+v_l^3 t^3}
  +\om^Q(t^2+\om^2 t+\om)^L\prod_{l=1}^{m_P}\frac{1+\om v_l
                        t}{1+v_l^3t^3}\right.}\ny\\ &&\hs\hs\hs\hs
+\;\left.\om^{2Q}(t^2+\om t+\om^2)^L\prod_{l=1}^{m_P}\frac{1+\om^2 v_l
t}{1+v_l^3t^3} \rb\hs\;\;\;\label{pol}   \EEA The integer $c$ is
chosen such that $\PP_Q$ becomes invariant against $t\ra\om t$ and $\PP_Q$ 
depends only on $s\equiv t^3$ with $\PP_Q(0)$ finite. Despite its denominators, 
$\PP_Q(s)$ will be a polynomial if the $v_l$ satisfy the Bethe-type equations:
\BEQ \lk\frac{1+\om\: v_j}{1+\om^2 v_j}\rk^L=(-1)^{m_P+1}\om^{Q-m_P}  
 \prod_{l=1}^{m_P}\lk\frac{v_l-\om\: v_j}{v_j-\om\: v_l}\rk. \EEQ
These Bethe-type equations resemble those of the spin-1 XXZ-chain.\\
As mentioned already above, the hamiltonian is obtained by expanding
the transfer matrix for small $p-q$ around a fixed value of $p$. 
If $\PP_Q$ has degree $m_E$, then from the $m_E$ zeros of (\ref{pol}) we get 
{\it a set} of $2^{m_E}$ energy eigenvalues
\BE E_i=a(1+\lam)+b\,-6\lk\pm w_1\pm w_2\pm\ldots\pm w_{m_E}\rk. 
\label{En} \EE where e.g. for the lowest sector $a=3m_E-2L,\:\;b=2Q$. 
Comparing to the formula eq.(\ref{EE}) obtained from the Onsager algebra, 
we see that $\;\;\cos{\theta_j}=(1+t_j^3)/(1-t_j^3).\;\;$ Our hamiltonian is 
hermitian
and so we should have $|\cos{\theta_j}|\le 1$ and all zeros $s_j=t_j^3$
must be on the negative real $s$-axis.\par 
A difficult question is whether all solutions to
these equations really give eigenvalues of our $H$, or whether there are
spurious solutions. This problem has been investigated in detail by
\cite{DKMC}.   
\subsection{Example: Sector $m_p=0$}
In general, in order to obtain energy eigenvalues of $H$ for chain length 
$L$ in the charge sector $Q$, one first has to choose a sector of given 
$m_P$ and to solve the Bethe-type equations
obtaining a set of Bethe-rapidities $v_l$. Then one has to use these $v_l$
to build up $\PP_Q$ and then to calculate its zeros $s_j$.
\par Obviously, the simplest case is the sector $m_P=0$ where there are no
$v_l$ and one can directly start with building the $\PP_Q$ and looking for 
the zeros $s_j$. E.g. for $Z_3$ with $L=20$ and $Q=1$ we calculate first 
\BEA \PP_1(s)&=&
20\,{s}^{13}+8455\,{s}^{12}+484500\,{s}^{11}+8533660\,{s}^{10}+  
61757600\,{s}^{9}\ny \\&+& 210859245\,{s}^{8}+363985680\,{s}^{7}
+326527350\,{s}^{6}+151419816\,{s}^{5}\ny \\&+&34880770\,{s}^{4}
+3656360\,{s}^{3}+146490\,{s}^{2}+1520\,s+1, \ny \EEA  then solve
numerically for its 13 zeros. Since the largest appearing coefficients
grow exponentially with $L$ (all coefficients are positive, and from 
(\ref{pol}) with $m_P=0$ we get $\PP_Q(1)=3^{L-1}$), this procedure becomes 
impractical for $L\la 
50.$ and e.g. energy eigenvalues for $L=1000$ can hardly be obtained 
this way. Can we get an analytic expression for the zeros of the
$\PP(s)$ for large {\it finite} $L$?
\par \cite{Baxgr} and \cite{AMCP} have obtained closed expressions for 
energy levels {\it in 
the limit $L\ra\infty$}. One method to get these is to rewrite the sum
over the terms determined by the zeros (chosing all $\pm$-signs to be
$+$ and so obtaining the lowest eigenvalue $E$) as an integral 
over a contour $C$ enclosing the negative real $t$-axis:
\BDM \sum_{j=1}^{m_E}\;\sqrt{1+2\lam\cos{\th_j}+\lam^2}
 =-\frac{3}{2\pi i}\int_{C}dt\;\sqrt{(1-\lam)^2+\frac{4\lam}{1-t^3}}
 \;\frac{d}{dt}\ln{\PP_Q(t)}    \EDM Then the contour is opened and gets
caught by three $120^o$ symmetrically oriented cuts, which arise 
from the square root of the integrand. 
The rightmost of these cuts is on the positive 
real axis at $\;\;1\;\le t\le |(1+\lam)/(1-\lam)|^{1/3}.\;\;\;$ 
All three cuts give the same contribution, and the rightmost one can be 
evaluated easily, since for $L\ra\infty$ only the term proportional 
$\;(t^2+t+1)^L\;$ of $\PP$, eq.(\ref{pol}), is non-oscillating and needs 
to be kept. Baxter (1988) finds that the ground-state energy 
for general $Z_N$ is given by (see also Albertini {\it et al.} 1989): 
\begin{equation} \lim_{L\ra\infty}\;\frac{E_0}{L}=
 -\frac{N}{\pi}\int_0^{\pi(1-N^{-1})}dx\:\sqrt{1-2\lam
  \frac{\sib^N(x+{\textstyle\frac{\pi}{N}})-\sib^N x}
 {\sib^N(x+{\textstyle\frac{\pi}{N}}) +\sib^N x}+\lam^2}.\hx\label{Bag}
 \end{equation}
\section{Determination of the zeros of the $\PP_Q(s)$}
We attempt to find the zeros of the $\PP_Q(s)$ for {\it finite L} on the 
negative real $s$-axis analytically. From the explicit expressions and 
numerical studies we observe:
\begin{itemize} \item[$\bullet$] For $L=Q\mo$ the zeros come in pairs 
which are reflection symmetric with respect to $s=-1$: with $s_j$ also 
$1/s_j$ is a zero. This remains approximately (within a shift) true 
 for the other charge sectors. \item[$\bullet$] The distances between 
adjacent zeros alternate slightly in size so that a smooth curve 
$s_j(k)$ is obtained only if we connect every second zero. 
  \end{itemize}
\subsection{Case $Z_3$, $m_P=0$} \noindent A useful variable transformation 
to be applied to $\PP_Q(s)$ turns out to be\\
$t=(u-\wud)/(u+\wud)\;\;$ or, writing $\;\;u=\tan{\beta}$: 
\BEQ t=\frac{\sin{(\beta-\pid)}}{\sin{(\beta+\pid)}}.  \EEQ
This maps $\;\;-\infty<s<0\;\;$ to e.g. $\;\;-\pid<\beta<\pid\;\;$ or 
$\;\;\frac{2\pi}{3}<\beta<\frac{4\pi}{3}\;\;$ (for definiteness let us 
choose the first interval), giving
\BEQ t^2+t+1=\frac{3}{\lk 2\sin{(\beta+\pid)}\rk^2}\hx\;\;\mbox{and}\hx\;\;
  t^2+\om t+\om^2=\frac{-3\,\om\:\cos{\beta}\:e^{-i\beta}}
   {2\lk\sin{(\beta+\pid)}\rk^2}. \EEQ so that
\BEQ \PP_Q\sim (t^2+t+1)^L+\om^Q(t^2+\om^2 t+\om)^L+\om^{-Q}(t^2+\om
t+\om^2)^L=0,\EEQ multiplying by the non-vanishing factors
$\lk\sin{(\beta+\pid)}\rk^{2L}$ becomes \\[-11mm]
\begin{center}
\begin{figure}[t]\vspace*{-7mm}
\psfig{file=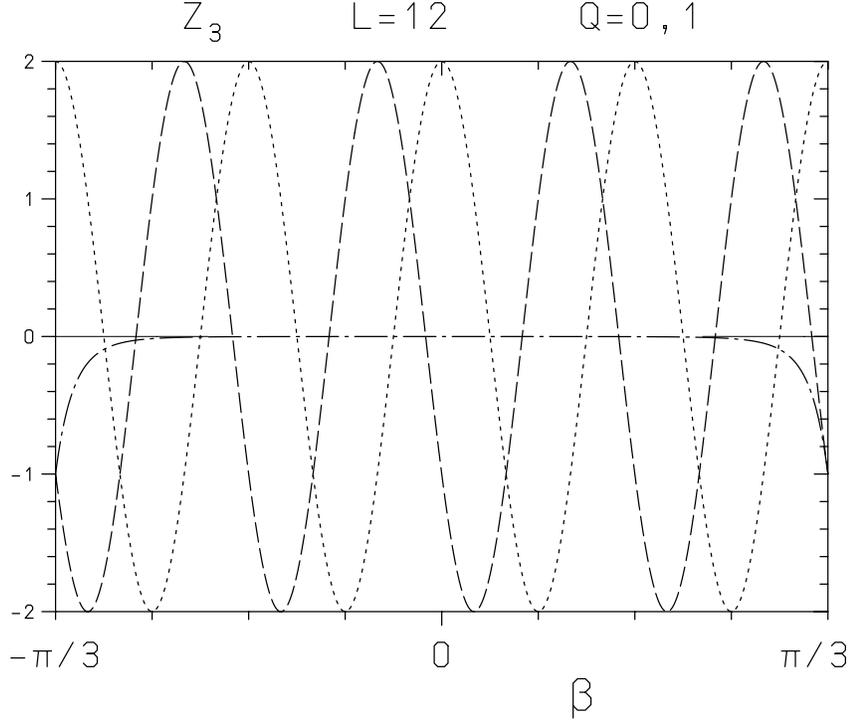,height=10.7cm}
\caption[ ]{Left-hand (dash-dotted line) and right hand (dotted and dashed) 
parts of (\ref{trig}) for $L=12$ sites. Charge sector $Q=0$: dotted lines, 
sector $Q=1$: dashed lines. The intersections with the dash-dotted line 
with $\;-\pi/3<\beta< \pi/3\;$ give the $m_E$ solutions $\beta_k$: $m_E=8$ 
for $Q=0$ and $m_E=7$ for $Q=1$.} \end{figure}\end{center} %
\begin{equation} -(2\cos{\beta})^{-L}\;=\;2\cos{\lk L\lb\beta+
 \frac{\pi}{3}\rb+\frac{2\pi Q}{3}\rk}.  \label{trig} \end{equation}
Now $\;\;-\frac{\pi}{3}<\beta<\frac{\pi}{3}\;\;$ means 
$\hx 2\cos{\beta}>1\hx$
so that for $L\gg 1$ the left hand side of (\ref{trig}) will be small, except
near the corners of the interval in $\beta$, as is shown in Fig. 1.
In order to get a solution, the fast oscillating right hand term has to 
coincide with the non oscillating left hand term. 
We see that all but the solutions near 
$\;\beta_k\approx-\frac{\pi}{3}\;$ and  $\;\beta_k\approx\frac{\pi}{3}\;$
are practically equally spaced in the variable $\beta$. 
Neglecting the left-hand side altogether, we get the approximate solutions
\BE \beta_k\approx K-\frac{\pi}{3};\hs K=\frac{6k+2Q-3}{6L}\,\pi;\hs
 k=1,2,\ldots,m_E. \label{zdn} \EE  or, expressed in terms of the
$\cos{\th_k}$ which we need for the energy eigenvalues:
\BE \cos{\th_k}\;\equiv\;\frac{1+t_k^3}{1-t_k^3}=-\frac{\sib^3(K+\pid)
 -\sib^3 K}{\sib^3(K+\pid)+\sib^3 K}. \label{adx}   \EE 
It is easy to see that using (\ref{zdn}, \ref{adx}) and replacing the 
summation in (\ref{En}) by an integral leads to Baxter's formula 
(\ref{Bag}) for the thermodynamic limit. Through a $Q$-dependent 
shift (\ref{zdn}) describes most of the $Q$-dependence of the finite-$L$ 
levels. Further finite-size effects are due to the deviations of the 
largest and smallest $\beta_k$ from the approximation (\ref{zdn}). 
\\ In order get an idea of the 
precision of (\ref{zdn}), consider e.g. $L=30,\;Q=0$. 
With $d_k=(\beta_k^{appr}/\beta_k^{exact})-1$ where $\beta_k^{appr}$ is 
the approximation (\ref{zdn}), we find 
\BDM d_1=0.0012;\;d_2=-1.89\cdot 10^{-5};\;d_3=6.48\cdot
10^{-7};\;d_4=-4.39\cdot 10^{-8};\;\ldots\EDM  For $L=30000$ and $Q=0$ 
we get $d_1=0.99\cdot 10^{-6}$, but compared to $L=30$ also the spacing 
between neighbouring $\beta_k$ is $\sim 2\cdot 1000/3$ times smaller.  
\par Taking the arcsin of (\ref{trig}) we get a pair of useful (still 
exact) equations for the $\beta_k$:
\BE  \mp\arcsin{\frac{1}{2\:(2\cos{\beta_k})^L}}\:+\:L\lk\beta_k
 +\frac{\pi}{3}\rk +\frac{2\pi Q}{3}\:-2\pi\,I_k^\pm
 \;\mp\frac{\pi}{2}=\;0.    \label{arc}  \EE 
The two sets of counting integers $I_k^\pm$ vary within the limits
\BEQ I_k^\pm\;=\;I_a^\pm,\;\;I_a^\pm\!+\!1,\ldots,\;I_b^\pm \EEQ
with  \BE  I_a^\pm\;=\;\Lb\frac{Q+3\mp 1}{3}\Rb;\hx\hx
   I_b^\pm\:=\:\Lb\frac{L+Q-1\mp 1}{3}\Rb; \label{cin} \EE The 
$\arcsin$-equation has the great advantage that from it each zero can be 
calculated separately by fixing $I_k^\pm$.
Numerically on a small PC we can easily get zeros with 40 digits precision
for up to $L=300000$ sites. Recall that computing directly the zeros of 
the $\PP_Q(t)$ we were limited to $L\la 50$. In addition, the arcsin-equation 
lends itself to excellent analytical approximations for 
the "corner"-solutions: Let us call $\xi$ the term which is important only
close to the corners:
\BE \xi\equiv \arcsin{ \frac{1}{2(2\cos{\beta})^L}} \label{cor} \EE
We find numerically that $\xi_k^\pm \ll 1$ 
even for the extreme cases $I_k^\pm = I_{a,b}^\pm$. We use this 
feature to approximate $\xi_k$. 
Let us consider the lower end of the interval in $\beta$, i.e. $I_k\ga I_a$. 
Then we get a behaviour smooth in $L$ when keeping $Q$ fixed (At the upper end
$I_k \la I_b$ the behaviour is smooth only if we keep $L+Q$ fixed mod 3).
Using  $\arcsin{x}\approx x$ and writing $\:\beta=-\pi/3+\ep\:$ we get 
(omitting the superscripts $\pm$): \BE \xi\;\approx\; 
\hal\,(2\,\cos{\beta})^{-L}=\hal\lk\cos{\ep}+\wud\sin{\ep}\rk^{-L}\;
\approx\; \hal\,\lk 1\,+\,\wud\,\ep\rk^{-L}. \label{be} \EE On the other 
hand, solving (\ref{arc}) for the $\beta_k$ which is not in $\xi_k$, 
we have:
\BE \ep_k=\frac{\Gamma_k \;\pm\;\xi_k}{L} 
 \hs\mbox{with}\hs \Gamma_k\,=\,\lk 2I_k\,-\frac{2Q}{3}\,
 \pm\frac{1}{2}\rk\,\pi. \label{epk} \EE 
Inserting (\ref{epk}) into (\ref{be}) and assuming $L\gg 1$
we get \BE \xi_k\approx \frac{1}{2}\lk 1+\frac{\sqrt{3}
  (\Gamma_k\:\pm\,\xi_k)}{L}\rk^{-L}\!\!\approx
    \frac{1}{2}\exp{\lb-\sqrt{3}(\Gamma_k\pm\xi_k)\rb} \label{nli}\EE 
This nonlinear equation for $\xi_k$, valid for small $\ep$, 
can be solved by iteration. The zeroth and first iterations give
 \BEQ  
\xi_k^{\pm(0)}=\,\frac{1}{2}\,e^{-\sqrt{3}\,\Gamma^\pm_k} \hx\;\;
\mbox{and}\hx\;\;
\xi_k^{\pm(1)}=\,\frac{1}{2}\,\exp{\lb-\sqrt{3}\lk\Gamma^\pm_k\: \pm
   \frac{1}{2}\:e^{-\sqrt{3}\:\Gamma^\pm_k}\rk\rb}, \EEQ
respectively, or  \BEQ 
 \beta_k^{\pm(1)}\:=\:-\frac{\pi}{3}\;+\;\frac{\Gamma_k^\pm}{L}
\;\pm\;\frac{1}{2L}\:\exp{\lb-\sqrt{3}\lk\Gamma_k^\pm \pm
   \frac{1}{2}\:e^{-\sqrt{3}\:\Gamma_k^\pm}\rk\rb},\hx\;\; 
etc.\label{bers}\EEQ  
So we find corrections to (\ref{zdn}) which fast become  
smaller with increasing $\;I_k$, and are useful mainly for two smallest 
values $\beta_k$. Closer to the middle of the interval, anyway 
the approximation $\xi_k=0$ gets very accurate. \\[1mm]
Coming back to the example $L=30000,\;Q=0$ given above, adding the $ith$
iteration corrections $\xi_k^{(i)}/L$ improves the approximation (\ref{zdn}) 
to \\$\hs\hx d_1^{(0)}=-5.48\cdot 10^{-8},\;\;d_1^{(1)}=3.285\cdot 
10^{-9},\;\;d_1^{(2)}=1.53\cdot 10^{-10}$ \\for the zeroth, first and second 
iteration, respectively. Regarding the $\beta_k$ away from the corners, 
already from (\ref{zdn}) we get e.g. $d_7=1.4\cdot 10^{-25}$, which 
improves slightly to $d_7^{(0)}=6.3\cdot 10^{-26}$ by adding the zeroth 
iteration. \par
Let us study how much the results for the $\xi_k$ are affected by the
approximations made in (\ref{be}) and (\ref{nli}), 
and how many iterations are needed. 
In Table 1 we give numerical values from our iteration formula in order
to compare them to the following numerical results from the exact formula 
(\ref{arc}).
A fit of the $L$-dependence of the numerical solution, obtained from
$L=25000,\ldots,\;125000$ sites is,
expressed in terms $\xi^\pm/\pi$:  {\small
\BEA Q=0+:&&+0.009927623657298\;+\;0.04835895/L\;+\;0.0315/L^2\;+\ldots
\ny\\Q=0-:&&-0.000045412321296\;-\;0.002017284/L\;-\;0.0338/L^2\;-\ldots
\ny\\Q=1+:&&+0.00000740160385053\;+\;0.0004910486/L\;+\;0.0130/L^2\;+\ldots
\ny\\Q=1-:&&-0.0017242062241336\;-\;0.023760499/L\;-\;0.0922/L^2\;-\ldots
\ny\\Q=2+:&&+0.0002780560828110\;+\;0.007462876/L\;+\;0.0685/L^2\;+\ldots
\ny\\Q=2-:&&-0.00000120676616449\;-\;0.000111825/L\;-\;0.00430/L^2\;-\ldots
 \ny\EEA} We see that for large $L$ our equation (\ref{nli}) 
determines the correction (\ref{cor}) to (\ref{zdn}) with excellent 
precision if its twice iterated solution is used. The coefficients in 
these expansions indicate that the asymptotic large-$L$-behaviour sets in 
only for about $L\ga 1000$. \par If we move downwards from the upper corner 
$\beta=\pi/3$, we get exactly the same numbers and formulae, just signs 
are changed and charge sectors are permuted. \\[-8mm]
\begin{table}[h]
\caption[]{Values of $\xi_k$ for the lowest value of $\beta_k$, as obtained 
from (\ref{nli}) at different iteration levels. We have multiplied the values
by  $10^6/\pi$ in order to avoid too small numbers.}
\begin{center}\begin{tabular}{|c|cc|cc|cc|}\hline
Iteration  & $Q=0\;+\;$ &$Q=0\;-\;$&$Q=1\;+\;$& $Q=1\;-\;$ &$Q=2\;+\;$&
$Q=2\;-\;$\\ \hline 0st   &10477.0&-45.401101&7.401901956&-1708.097&
 278.47707&-1.20675824029\\
     1st& 9896.39&-45.412318&7.401603838&-1724.047&   
 278.05541&-1.20676616444\\
     2nd & 9927.70&-45.412321&7.401603850&-1724.196&
 278.05605&-1.20676616449\\ \hline 
\end{tabular} \end{center}\end{table}\\[-8mm]
\subsection{The leading finite-size corrections}
The formulae of the last section are sufficient to calculate
finite-size eigenvalues. But often it is convenient to transform the
sums in (\ref{En}) into integrals in order to have a compact expression also
for finite $L$ eigenvalues or to get the leading finite-size corrections to 
the thermodynamic limit results. This can be done e.g. applying  
Euler-Maclaurin-techniques as is common in calculations for the XXZ-quantum
chain. Following \cite{WEk} and \cite{Ham}, but taking care of the 
alternating spacing of the zeros, we define two functions $Z_L^\pm$
\BEQ Z_L^{\pm}(\beta)\;=\: 
 \frac{v_L^\pm}{2\pi\,L}\lb\mp\arcsin{\frac{1}{2\:(2\cos{\beta})^L}}\:
 +\:L\lk\beta+\frac{\pi}{3}\rk +\frac{2\pi
Q}{3}\:\mp\frac{\pi}{2}\rb,\EEQ where $v_L^\pm\,=\,3m_E^\pm/L$ so that
 $\lim_{L\ra\infty}v_L^\pm\:=\:1\;$. From (\ref{cin}) we read off
$\;m_E^\pm=I_b^\pm-I_a^\pm+1.\;$
In $Z_L^\pm$ the roots $\beta_k$ are exactly equidistant.
We define the root density \BEQ \sig^\pm_L(\beta)\;=\;dZ_L^{\pm}/d\beta
   \;=\;\frac{v_L^\pm}{2\pi}\lb 1\:\mp\:(-1)^L\:
\frac{\tan{\beta}}{\sqrt{4\,(2\cos{\beta})^{2L}\:-1}}\rb,\EEQ
and use $\;\; \sig_\infty^+(\beta)+\sig_\infty^-(\beta)
   =\sig_\infty,\;\;\;m_E^++m_E^-=m_E.\:$  
Then the finite-size correction to a quantity $\;\;\int 
 d\beta\; \sig_\infty\;w(\beta)\;\;$ (since we are mainly interested in 
the ground-state energy per site, this will be the $w$ of (\ref{En})) 
$\;\;$ is \BEA \lefteqn{\frac{1}{L}\lk\sum_{k=1}^{m_E^+}\:w(\beta_k^+)
+\sum_{k=1}^{m_E^-}w(\beta_k^-)\rk\;-\; \int_{-\pi/3}^{\pi/3}d\beta\;
  \sig_\infty\;w(\beta)}\ny \\ &=&
\int_{-\pi/3}^{\pi/3}d\beta\;
  (\sig^+_L+\sig^-_L-\sig_\infty)\;w(\beta)\;
  -\sum_\pm \lk\int_{-\pi/3}^{\beta_a^\pm}+
 \int_{\beta_b^\pm}^{\pi/3}\rk\:d\beta\;\sig_L^\pm\;w(\beta)\hx\;\;\;
 \ny\\&+&\frac{1}{2L}\sum_\pm\lb
 w(\beta_b^\pm)+w(\beta_a^\pm)\rb
  +\;\frac{1}{12 L^2}\sum_\pm\lk
\frac{w'(\beta_b^\pm)}{\sig_L^\pm(\beta_b^\pm)}-
  \frac{w'(\beta_a^\pm)}{\sig_L^\pm(\beta_a^\pm)}\rk\;+\ldots\;\; 
  \label{fsc} \EEA
Both $\; \beta^\pm_a\equiv\beta_1^\pm=\beta^\pm(I_a^\pm)\;$ and 
$\;\beta^\pm_b=\beta^\pm(I^\pm_b)\;$ depend on the charge $Q$.  
\section{The zeros for higher $Z_N$-superintegrable cases}
Analogous trigonometic formulae result for the higher $Z_N$-cases. With
increasing $N$, the equations get more involved, but have a similar 
structure.    
\subsection{The case $Z_4$:} 
Since we transform the variable $t$ and not $s=t^4$, in order to
map the negative real $s$-axis to a range of real $\beta$, we have to
include a phase and use
\BE t=e^{i\pi/4}\;\frac{\sin{(\beta-\pivi)}}{\sin{(\beta+\pivi)}}
\label{vi} \EE
so that $-\infty<s<0$ is mapped to e.g. $-\pivi<\beta<\pivi$. After some 
algebra, we get that the condition for the zeros of the 
$m_P=0$-functions $\PP_Q(s)$
\BEA \lefteqn{ \PP_Q(s)=t^{-c}\lb
(t^3+t^2+t+1)^L+(-1)^Q(t^3-t^2+t-1)^L \right.}\ny\\ 
 &&\left.\hs + i^{-Q}(t^3+it^2-t-i)^L+i^Q(t^3-it^2-t+i)^L\rb\;=\;0 \EEA
translates to\\[-3mm]  
\BEA \lk\!\frac{\sqrt{2}-\cos{(2\beta)}}{\sqrt{2}+
  \cos{(2\beta)}}\!\rk^{L/2}\!\!\!
  \cos{\lb\! L\lk\rho_-\!+\frac{\pi}{8}\rk +\frac{\pi Q}{4}\rb}
  +\cos{\lb  L\lk\rho_++\frac{3\pi}{8}\rk
  +\frac{3\pi Q}{4}\rb} \ny \\ \hspace{2cm} =0\hs\hs  \label{avi} \EEA
where $\hs\hs\rho_\pm= \arctan{\lb\lk\sqrt{2}\pm 1\rk\tan{\beta}\rb}.$
\\[3mm]
For $L\gg 1$ and $\cos{(2\beta)}$ not too 
close to zero, the second term of (\ref{avi}) dominates and the solutions 
come where   \BEQ \cos{\lb L(\rho_+ +{\textstyle\frac{3\pi}{8}})\:
  +{\textstyle\frac{3\pi Q}{4}}\rb}\;\approx\;0.\EEQ 
After some rearrangement this leads to
\BE \tan{\beta_k}\approx-\tan{\frac{\pi}{4}}\tan{\frac{\pi}{8}}
   \cot{\lk K\,+\frac{\pi}{8}\rk}
 \hx\;\;\mbox{with}\hx\;\;K\:=\,\frac{(4k+Q-2)\pi}{4L},\label{bK}\EE  
$k=1,2,\ldots,[(3L-Q)/4]\;\;$ and e.g. 
$-\pivi<\beta_k<\pivi$. In contrast to the $Z_3$-case, where in the 
approximation (\ref{zdn}) the zeros were equally spaced in $\beta$, here 
this is no longer the case: now the $\beta_k$ are more dense in the center 
region. In the center $\beta\approx 0$ the oscillations of the 
$\rho_+$-term in (\ref{avi}) are 
$(\sqrt{2}+1)/(\sqrt{2}-1)\approx 5.83$ times faster than those of the
$\rho_-$-term.  (\ref{vi}) and (\ref{bK}) are equivalent to
\BEQ  \cos{\theta}_k\;\equiv\;\frac{1+t_k^4}{1-t_k^4}\;\approx\;
 -\frac{\sib^4(K+\pivi)-\sib^4(K)}{\sib^4(K+\pivi)+\sib^4(K)}.\EEQ
As in the $Z_3$-case, in the thermodynamic limit, this approximation 
leads back to Baxter's ground-state energy formula. 
\subsection{The case $Z_5$:}
Transforming
$\;\; t\:=\:\sin{(\beta-\epf)}/\sin{(\beta+\epf)}$, the condition
\BEQ \PP_Q\sim \JJ_0^L+\om^{-Q}\JJ_1^L+\om^Q\JJ_4^L+\om^{-2Q}\JJ_2^L
 +\om^{2Q}\JJ_3^L=0 \EEQ  with $\;\; \JJ_k=t^4+\om^k 
t^3+\om^{2k} t^2 +\om^{3k} t+\om^{4k} \;\;$  takes the form
\BEA 
\lefteqn{\frac{1}{2}\:\lk\frac{\cos{\beta}}{\cos{\frac{\pi}{5}}}\rk^{-L}
 +\:\cos{\lb L\lk\beta+\frac{\pi}{5}\rk+\frac{2\pi Q}{5}\rb}}
 \ny \\ &&\hs\hs\hs+\: B^{-L/2}\cos{\lb L\lk\chi+\zpf\rk
 +\frac{4\pi Q}{5}\rb}\:=\:0.\hs\label{trif}\EEA where ${\displaystyle
\hx\;\; \tan{\chi}=\cot{\frac{\pi}{5}}\,\cot{\frac{\pi}{10}}\,
\tan{\beta}\hx\;\;\mbox{and}\hx\;\; 
 B\,=\frac{4\,\sbq +\sqrt{5}-2}{2\,\cos{\tepf}}}$.\\[3mm]   
The solutions come for $-\infty<s<0$ corresponding to e.g. $-\epf<\beta<\epf$.
Since for $L\gg 1$ and $|\beta|$ not too close to $\epf$ we have 
$1\ll (\cos{\beta}/\cos{{\textstyle \epf}})^{-L}\ll B^{-L/2},$ 
the solutions come close to where the cosine containing $\chi$ vanishes, 
i.e. at \BEQ\tan{\beta_k}\:\approx\:-\tan{\frac{\pi}{5}}\tan{\frac{\pi}{10}}
   \cot{\lk K\,+\frac{\pi}{10}\rk}; \hs K=\frac{(10k+2Q-5)\pi}{10L}
\label{beap}\EEQ and integers $k=1,2,\ldots$ such that e.g. $|\beta_k|<\epf$. 
Equivalently, we can write
\BE \cos{\theta_k}\equiv\frac{1+t_k^5}{1-t_k^5}\:
 \approx\:-\frac{\sib^5(K+\epf)
 -\sib^5 K}{\sib^5(K+\epf)+\sib^5 K}. \label{cK}\EE
The $Z_3$-formula (\ref{trig}) has the same form as the first terms of 
(\ref{trif}) since $2\cos{\beta}=\cos{\beta}/\cos{\frac{\pi}{3}}.$
The expressions for general $Z_N$ will be given elsewhere.
\subsection{The case $Z_6$:}
We conclude with the case $Z_6$: As in the case of $Z_4$, because $N=6$ is 
even, we include 
a phase and transform $\;\;-\infty<s<0\;\;$ to $\;\;-\pis<\beta<\pis\;\;\;$
using \BEQ s\,=\,t^6;\hs\hs 
t=e^{i\pi/6}\;\frac{\sin{(\beta-\pis)}}{\sin{(\beta+\pis)}}.  \EEQ
Now $\;\;\PP_Q(s)=0\;\;$ (we don't give the explicit $Z_6$ formula for 
$\PP_Q$, since this is easily generalized from the $Z_5$-case) after 
some algebra leads to   
\BEA \lefteqn{\lk\frac{1-\al}{1+\al}\rk^{L/2}\!\!
 \cos{\lb L\lk\rho_- +\!\frac{\pi}{12}\rk+\frac{\pi Q}{6}\rb} 
  +\lk 1-\al\rk^{L/2}\cos{\lb L\lk\rho_0 +\!\frac{\pi}{4}\rk
  +\frac{\pi Q}{2}\rb}}\ny\\&&\hspace{46mm}
 +\;\cos{\lb L\lk\rho_+ +\!\frac{5\pi}{12}\rk+\frac{5\pi Q}{6}\rb}
  \;=\;0\hs\hx  \label{zse} \EEA\\[-7mm]  
\BEA \lefteqn{\mbox{with}\hs\hx \al=\frac{\sqrt{3}}{2}\;
 \frac{\tib\pis-\tib\beta}{\tib\pis+\tib\beta}\hs\mbox{and}}\ny \\
  &&\rho_0=\arctan{\lk\sqrt{3}\,\tan{\beta}\rk};\hs
 \rho_\pm\:=\:\arctan{\lb\lk 2\pm\sqrt{3}\rk\sqrt{3}\,\tan{\beta}\rb}.\hs 
\EEA For $L\gg 1\;$ and $\;\al\;$ not too close to zero, the last term 
containing $\;L\rho_+\;$ dominates and the solutions $\:\beta_k\:$ approach  
\BEQ  \tan{\beta_k}\:\approx\:-\tan{\frac{\pi}{6}}\tan{\frac{\pi}{12}}
   \cot{\lk K\,+\frac{\pi}{12}\rk};\hs K=\frac{(6k+Q-3)\pi}{6\:L} \EEQ 
and $k=1,2,\ldots,m_E$ with $m_E=[(5L-Q)/6].$  
We can rewrite this like (\ref{cK}).\\[2mm] 
\section{Conclusions}
In the analytic solution of the superintegrable chiral $Z_N$-Potts quantum
chain the energy eigenvalues are determined by the zeros of special 
polynomials. These polynomials also play a central role in the representation
theory of the Onsager algebra. In the thermodynamic limit, sums over
functions of the zeros can be evaluated by contour integrations. 
For finite systems we can try to calculate the zeros numerically. However, 
applying standard computer routines directly to the explicit form of 
the polynomials will run into trouble beyond some chain length $L$,  
because the largest coefficients of the polynomials are growing exponentially 
with increasing $L$. So we were looking for an 
alternative and possibly analytic approach. \par
By a suitable change of variable we transform the equations for the zeros 
into trigonometric equations, for which approximate analytic solutions
can be obtained. We give explicit formulae for the simplest 
polynomials.\par\noindent 
In the case of $Z_3$ the solutions corresponding to the zeros of the 
polynomials are almost equidistant in the variable $\beta$. However, 
for $Z_4$ and higher, 
the density of zeros in $\beta$ varies appreciably over the finite 
interval. Our formula describes the solutions, except those near at the
boundaries, with exponential precision in $L$. 
The leading finite-size effects are due to the solutions at the 
extreme ends of the interval in $\beta$.
For the case $Z_3$ we derive a special formula which describes also the 
extreme solutions, thus allowing to calculate finite-size energy 
eigenvalues for the low-lying sectors.    
\subsection*{Acknowledgements}
I am grateful to Roger E. Behrend for many fruitful discussions.\\ 
Part of this work was performed during a visit of the author to SISSA, 
Trieste where it was supported by the TIMR-Network Contract FMRX-CT96-0012 of 
the European Commission. I thank G. Mussardo for his kind hospitality at 
SISSA.   %

\end{document}